\documentstyle[fleqn,espcrc2,epsf,epsfig]{article}

\begin{document}
\title{
Prospects of `Topologically Unquenched QCD' from a study of the analogous
importance sampling method in the massive Schwinger model
\thanks{Supported by DOE contract DE-FG03-96ER40956
and in part by the Swiss National Science Foundation (SNF).}
}
\author{S. D\"urr
\address{University of Washington, Physics Department, Box 351560,
Seattle, WA 98195, U.S.A.}
}
\begin{abstract}
I give a quick summary of my proposal for simulating an improvement on quenched
QCD with dynamical fermions which interact with the gluon configuration only
via the topological index of the latter. It amounts to include only the
topological part of the functional determinant into the measure, thereby
absorbing a correction factor into the observable. I discuss the prospects of
this concept from a study in the massive $N_f$-flavour Schwinger model, where
the correction factor is indeed found to be of order $O(1)$.
\end{abstract}
\maketitle

Simulations of lattice QCD have, to date, used the quenched and the partially
quenched approximations, where internal fermion loops are omitted or heavily
suppressed by giving the dynamical quarks a mass heavier than that of the
propagating current quarks.
Such an approximation turns out to be necessary, since presently available
update algorithms with dynamical fermions tend to slow down rather dramatically
if the fermion mass is taken light.
From a physical point of view this critical slowing down does not come as a
surprise, since it reflects the very nonlocal nature of the fermion determinant.
The problem with the quenched and, to a milder extent, the partially quenched
approximations, however, is that they introduce unphysical degrees of freedom
which persist in low energy observables and render the whole theory sick, if
the chiral limit is taken \cite{QuenchedQCD,PartiallyQuenchedQCD}.

An alternative path towards full QCD is ``To\-pologically unquenched QCD''
\cite{TUQCD}. The aim is to simulate full QCD, but in an unconventional way:
The functional determinant is split into a ``topological factor'', which is
included into the measure and a ``correction factor'' which is absorbed in
the observable.
The ``topological factor'' is defined as the determinant of a suitably chosen
``standard configuration'' which shares with the actual configuration nothing
but the overall topological charge.
This means that the importance sampling pays attention only to the first
(``topological'') factor in the decomposition
\begin{equation}
\det(D\!\!\!\!/\,+m)=
\det(D\!\!\!\!/\,_{\rm std}\!+m)\,
{\det(D\!\!\!\!/\,+m)\over\det(D\!\!\!\!/\,_{\rm std}\!+m)}
\,.
\end{equation}
In the language of full QCD this amounts to a statistical guess of the
determinant of a configuration on the basis of its topological index.
The last factor (i.e. the ratio of the true determinant of the configuration
to the guess on the basis of which it has gotten accepted in the
``topologically unquenched'' Metropolis test) is calculated only when a
measurement is done (into which it is included).
Hence, the number of determinant ratios to be calculated is reduced quite
drastically -- at the price of computing for a configuration, whenever possible
(see below), its topological index.

From this outline it is clear that for the concept to work two conditions have
to hold true.
First, there must be a correlation between the logarithm of the determinant
evaluated on a certain configuration and the topological index of the latter.
Second, the costs (in terms of CPU time) for determining the topological index
must be smaller than the savings on determinant evaluation.

The idea of ``topologically unquenched QCD'' has been launched \cite{TUQCD}
with a strong focus on theoretical aspects, merely arguing that a correlation
between the functional determinant and the topological index (in the continuum
theory) is to be expected from the Instanton Liquid Model \cite{SchaferShuryak}.
At this time is seems appropriate to do the next step and to put it to the
(numerical) test.
A toy theory where the principal issue of ``topological unquenching''
can be studied without having to worry about many technical aspects of
the implementation (which do indeed matter in QCD) is the multi-flavour
Schwinger model (MSM), i.e. QED(2) with $N_f\geq2$ light degenerate fermions.
The physics of this theory resembles in many aspects QCD(4) {\em slightly
above\/} the phase transition (see refs. \cite{SchwingerModel} and references
cited in the second of them).
Technically, however, there is a difference: Even though the gauge fields in
the MSM (in the continuum-version on the torus) fall into topologically
distinct classes, there is no precise analog of the instanton, i.e.$\,$no
topologically nontrivial object which is simultaneously a minimum of the
classical action and localized in space-time.
As a consequence, the ``standard configuration'' in each topological sector
cannot be constructed artificially but must be chosen as the ``most typical''
configuration in that sector (i.e. the one \footnote{
In practice, it proves useful to relax this condition somewhat and to construct
the ``standard determinant'' as the geometric average of a few determinants
of configurations which have an action close to the sectoral average.}
which is in its classical/effective action closest to the sectoral average) ---
cf. strategy $(ii)$ in \cite{TUQCD}.

In order to test the ``topological unquenching'' idea, I have chosen to
implement the Schwinger model with 2 dynamical staggered fermions both full
and ``topologically unquenched'', as well as the quenched approximation,
always using the unimproved gauge action
$S_{\rm class}\!=\!\beta\sum(1\!-\cos(\theta_\Box\!))$.
The three theories are compared to each other using each time $\sim 1500$
independent indexed (see below) configurations on a lattice of common size
$V\!=\!14\times 6$ with periodic/thermal boundary conditions.
The common coupling is $\beta\!=\!1/g^2\!=\!3.0$ and the fermion
mass is $m\!=\!0.1$.
These values are chosen such that the ``pion'' (pseudo-scalar iso-triplet)
has a mass $M_\pi\!\simeq\!2.066\;g^{1/3}\;m^{2/3}\!\simeq\!0.3706$ as to fit
into the box. The formula used is the prediction by the bosonized
(strong-coupling) version of the theory, which seems adequate, since
$m/g\simeq 0.1732\ll 1$.
Note that, unlike in QCD, the power by which $M_\pi$ depends on $m$ is
specific for $N_f\!\!=\!2$.
Since the operator $D\!\!\!\!/\,+m$ is represented by a small ($84\times 84$)
matrix with 5 entries per row or column, I have decided to compute the
determinants exactly, using the routines CGEFA and CGEDI from the LINPACK
package.
For the topological charge, I have implemented both the L\"uscher
geometric definition $\nu_{\rm geo}\!=\!{1\over2\pi}\sum\log(U_\Box)$ and the
rescaled naive $\nu_{\rm nai}\!=\!\kappa_{\rm nai}\sum\sin(\theta_\Box)$ with
$\kappa_{\rm nai}\!=\!1/(1-\langle S_{\rm class}/V\rangle)$.
A configuration is assigned an index and used for measurements only if
the geometric and the naive definition, after rounding to the nearest integer,
agree (which, at $\beta\!=\!3.0$, holds true for $\sim 93\%$ of them).

\begin{figure}[t]
\epsfig{file=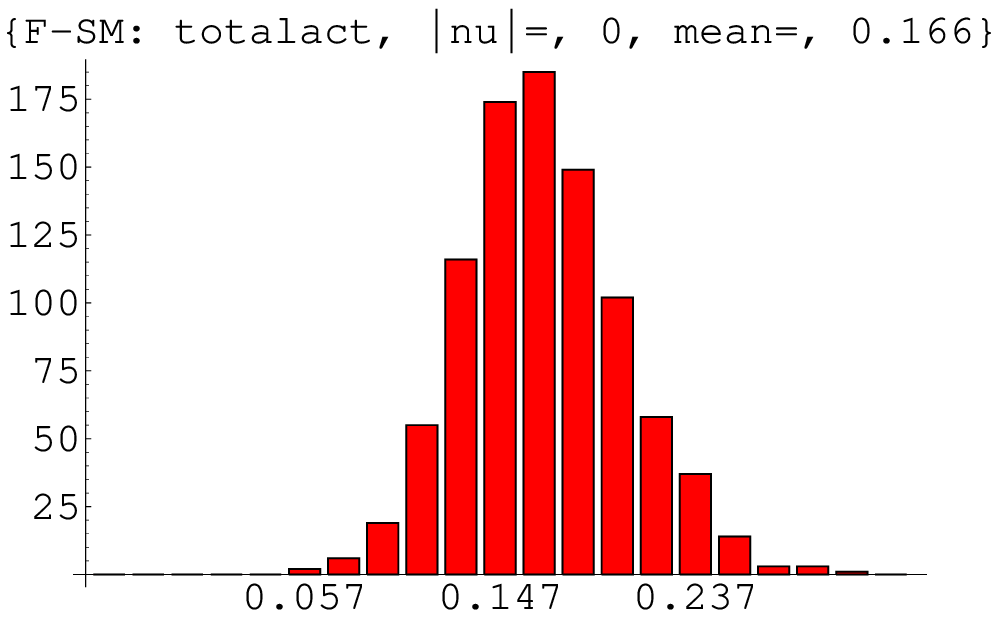,height=2.4cm,width=3.2cm,angle=90}
\hspace*{-0.1cm}
\epsfig{file=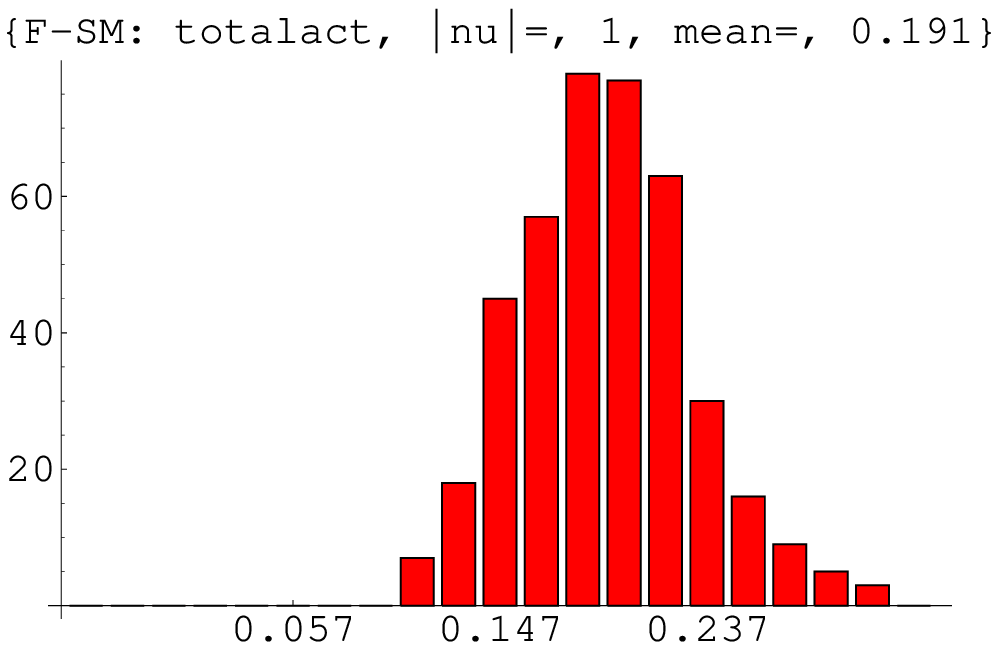,height=2.4cm,width=3.2cm,angle=90}
\hspace*{-0.1cm}
\epsfig{file=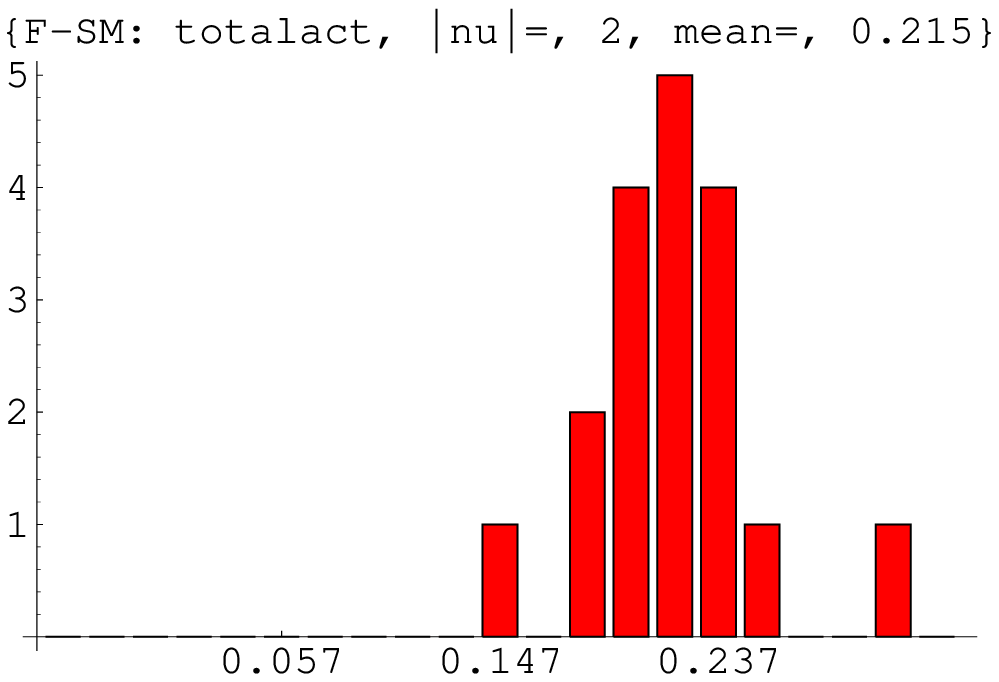,height=2.4cm,width=3.2cm,angle=90}
\epsfig{file=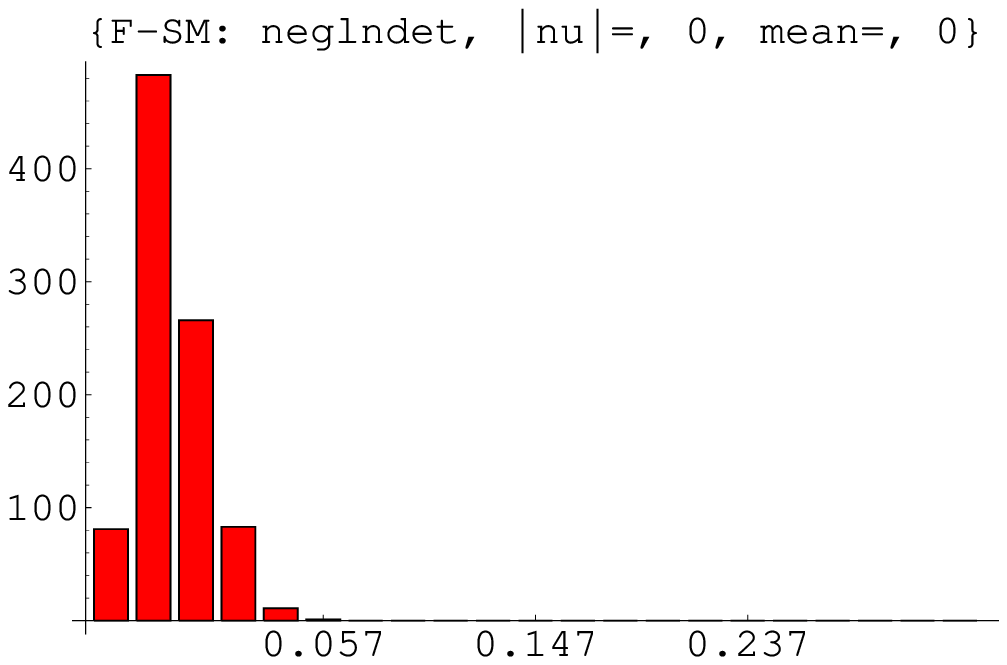,height=2.4cm,width=3.2cm,angle=90}
\hspace*{-0.1cm}
\epsfig{file=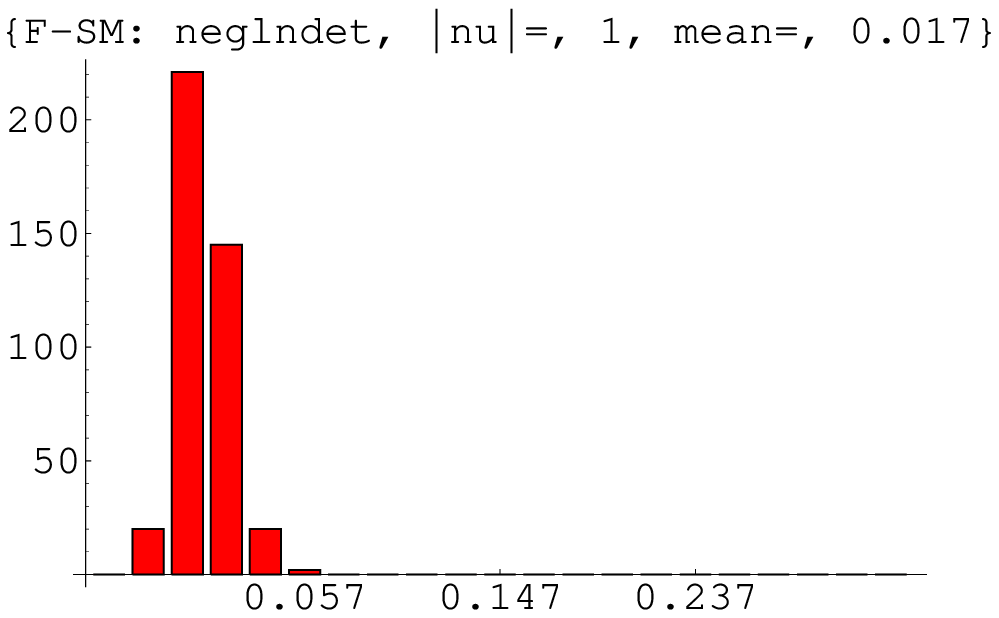,height=2.4cm,width=3.2cm,angle=90}
\hspace*{-0.1cm}
\epsfig{file=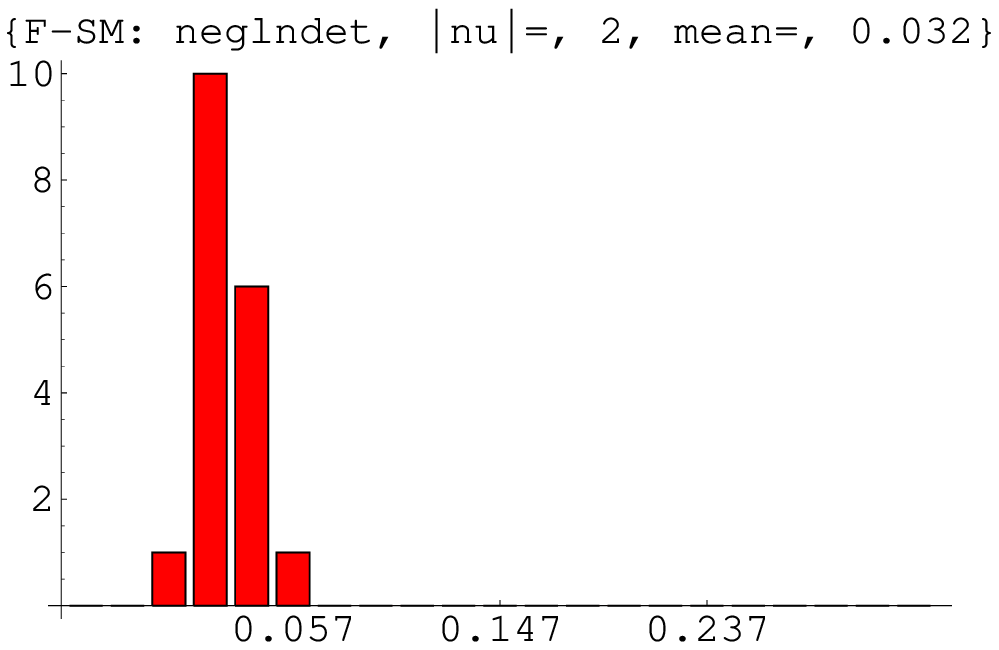,height=2.4cm,width=3.2cm,angle=90}
\epsfig{file=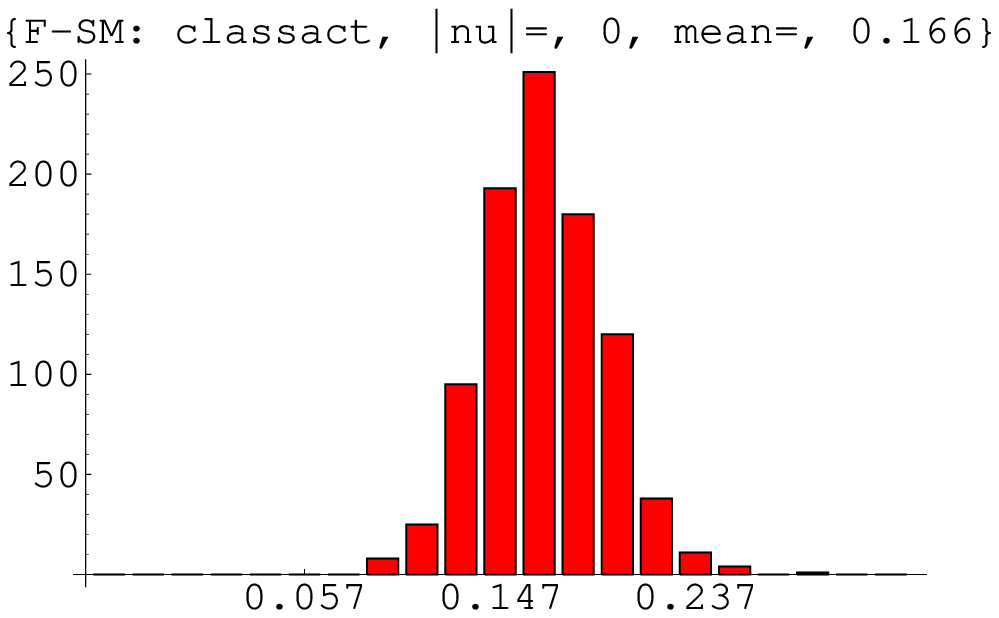,height=2.4cm,width=3.2cm,angle=90}
\hspace*{-0.1cm}
\epsfig{file=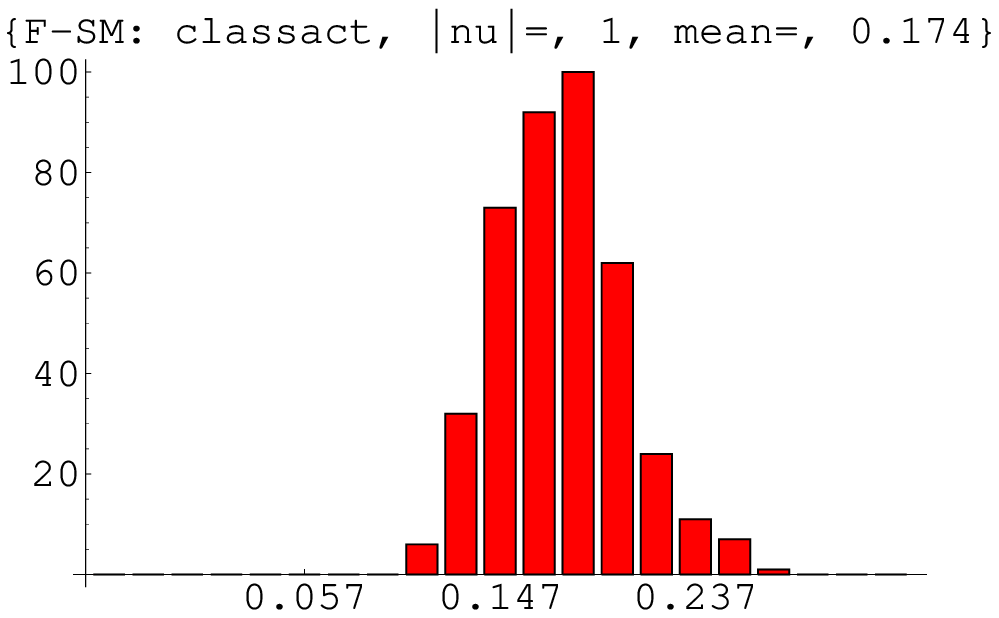,height=2.4cm,width=3.2cm,angle=90}
\hspace*{-0.1cm}
\epsfig{file=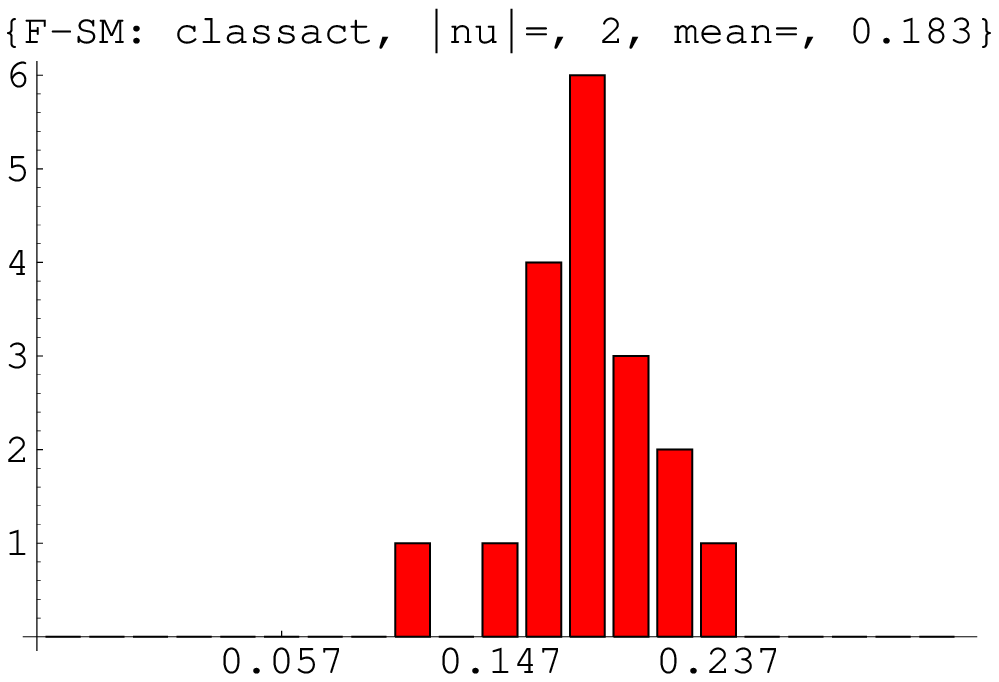,height=2.4cm,width=3.2cm,angle=90}
\vspace*{-1.2cm}
\caption{\sl Histograms of $S_{\rm class}$, $-\log(\det(D\!\!\!\!/\,+m))$
[for 2 continuum flavours, after an overall-shift] and $S_{\rm tot}$ in the
full theory with a pair of staggered fermions ($\beta\!=\!3.0, m\!=\!0.1$) --
each for $|\nu|\!=\!0,1,2$.}
\vspace*{-0.6cm}
\end{figure}

The first thing to look at is whether there is a correlation between
the effective action or its components ($S_{\rm class},
-\log(\det(D\!\!\!\!/\,+m))$) of a configuration and its topological index.
The result which merely extends work initiated in the literature
\cite{Gattringeretal} is shown in fig.$\,$1.
Only the center of the distribution of $-\log(\det(D\!\!\!\!/\,+m))$
(normalized per plaquette) is substantially shifted upwards when $|\nu|$
increases. The shift between two neighboring sectors is about half of the
width of the distribution in each of them.
Thus there is a positive correlation between $-\log(\det(D\!\!\!\!/\,+m))$
and $|\nu|$, and the ``topological unquenching'' idea might work.

\begin{figure}[t]
\epsfig{file=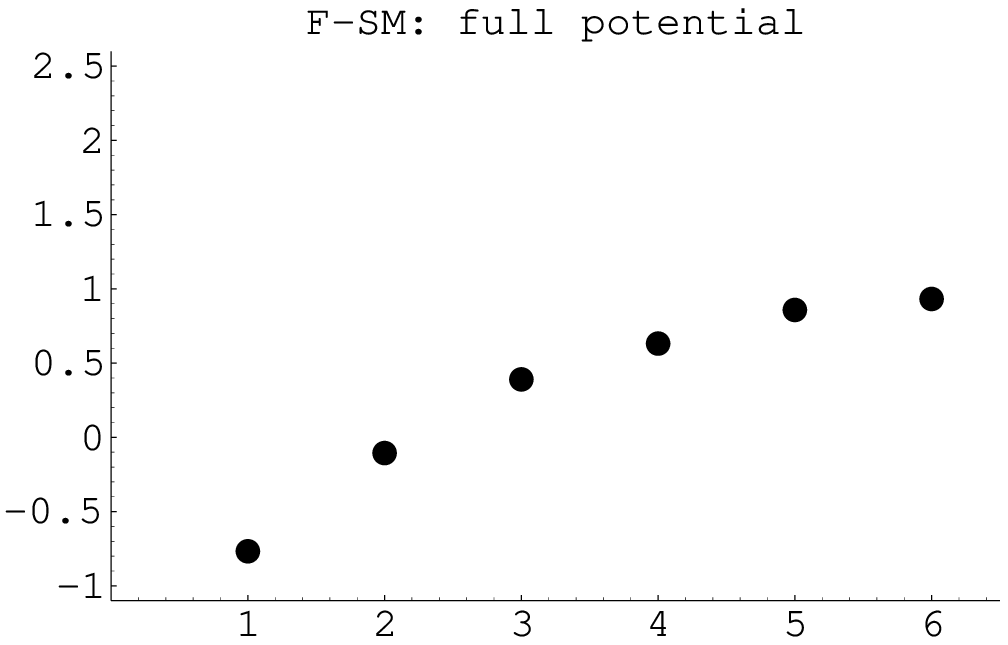,height=2.2cm,width=3.6cm,angle=0}
\hspace*{-0.1cm}
\epsfig{file=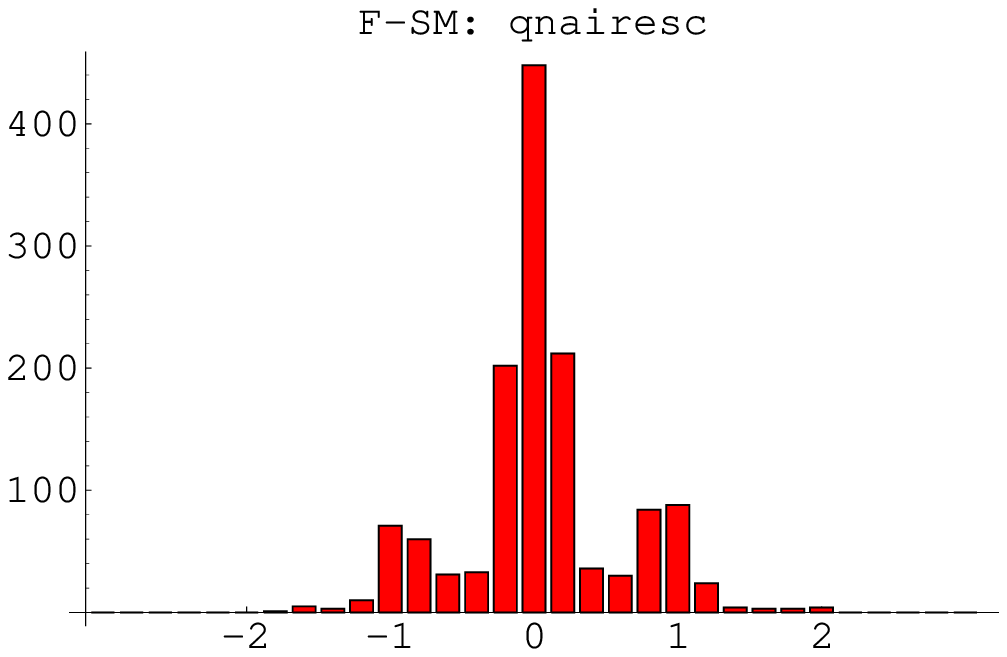,height=2.2cm,width=3.6cm,angle=0}
\epsfig{file=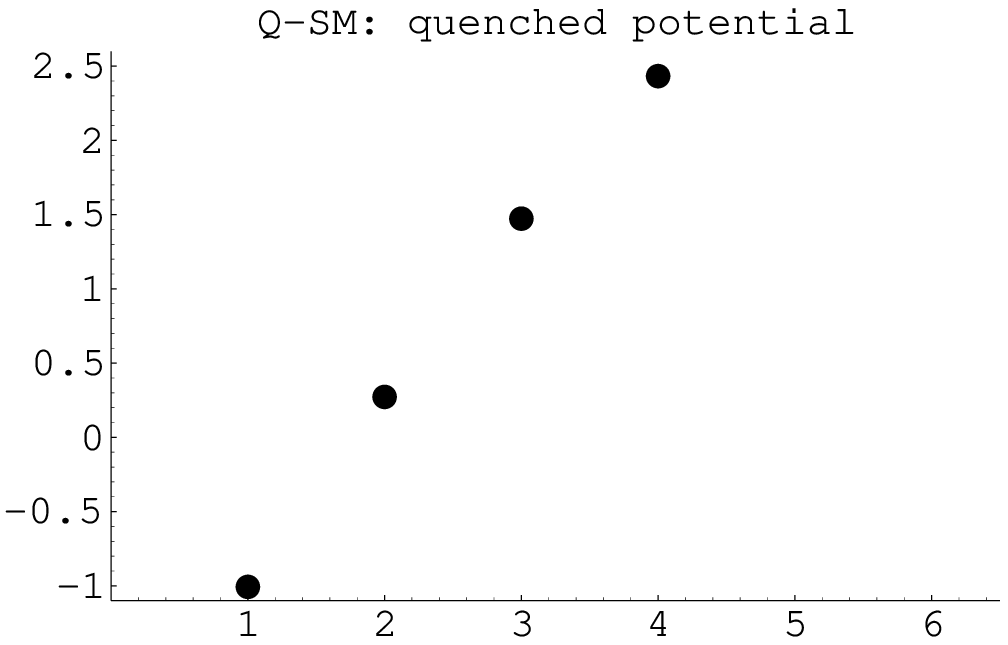,height=2.2cm,width=3.6cm,angle=0}
\hspace*{-0.1cm}
\epsfig{file=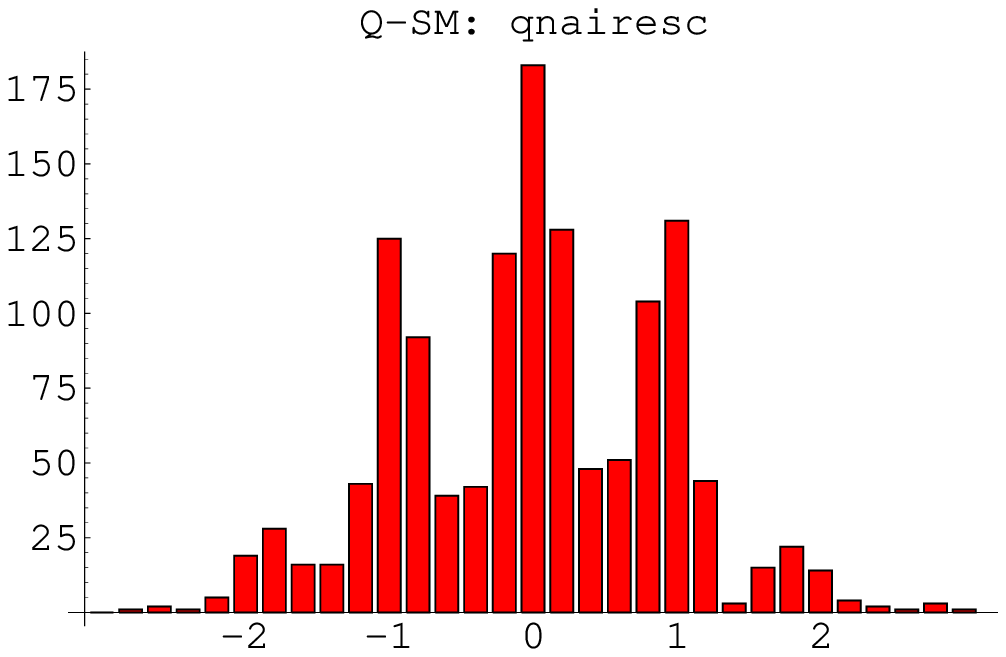,height=2.2cm,width=3.6cm,angle=0}
\epsfig{file=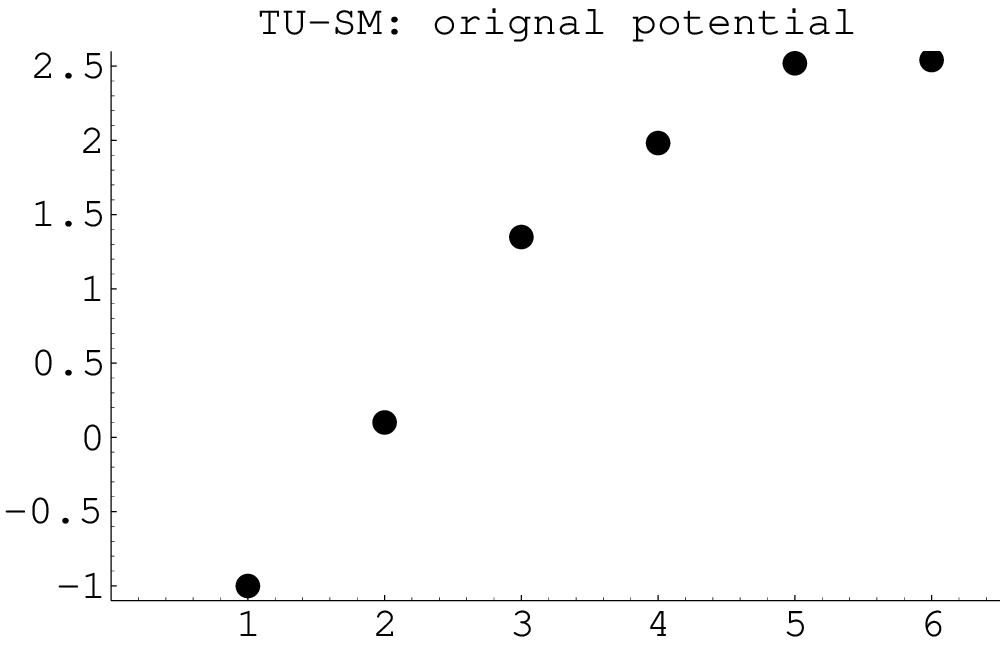,height=2.2cm,width=3.6cm,angle=0}
\hspace*{-0.1cm}
\epsfig{file=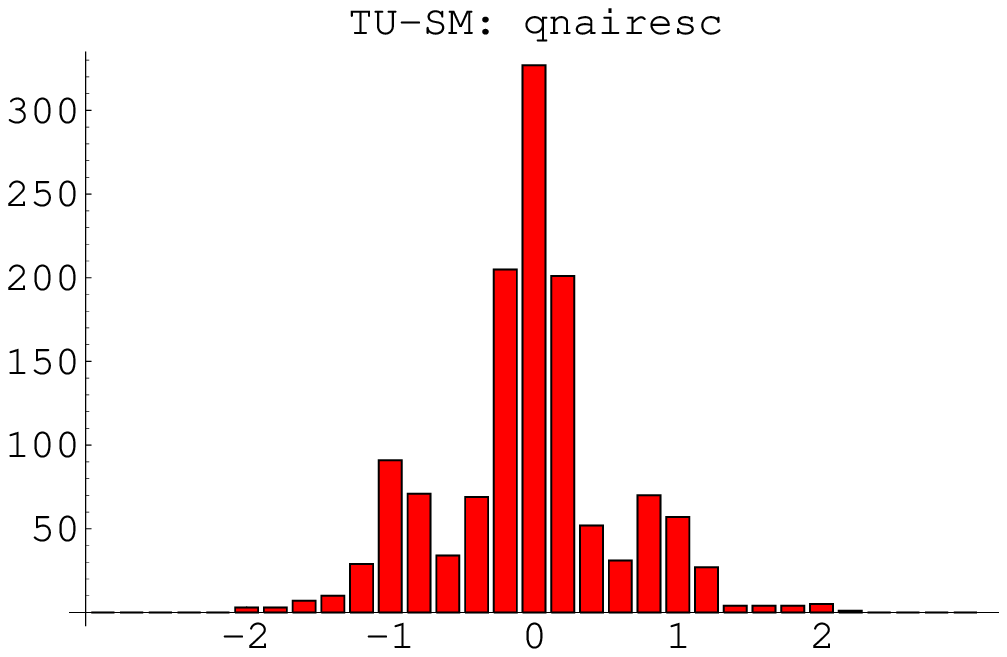,height=2.2cm,width=3.6cm,angle=0}
\epsfig{file=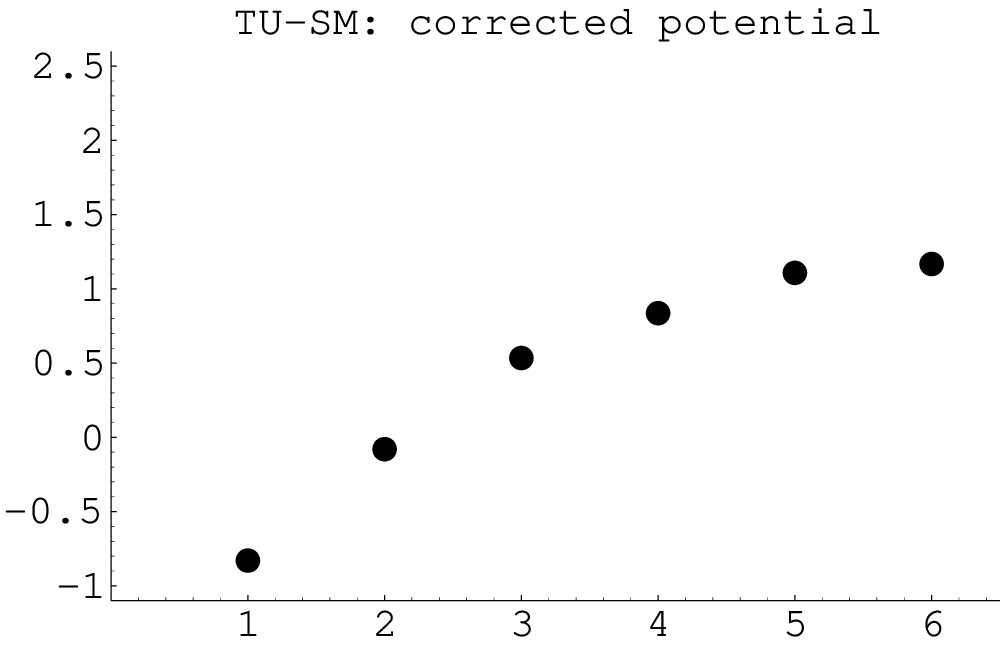,height=2.2cm,width=3.6cm,angle=0}
\hspace*{-0.1cm}
\epsfig{file=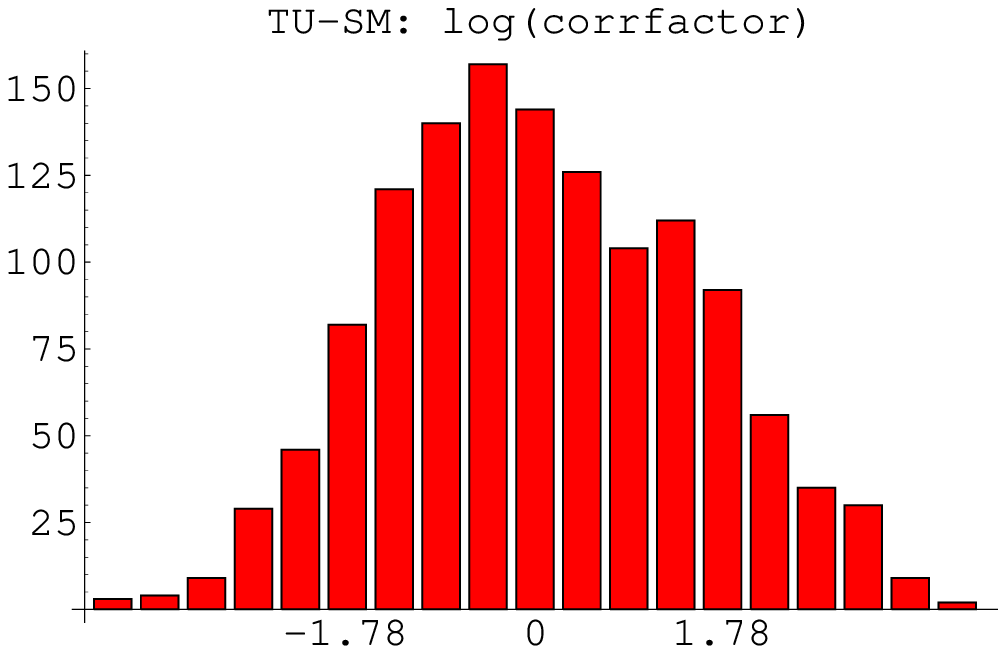,height=2.2cm,width=3.6cm,angle=0}
\vspace*{-0.8cm}
\caption{\sl Static quark-antiquark potential and distribution of
$\nu_{\rm nai}$ in the full, quenched and ``topologically unquenched''
(without the correction factor) theories, as well as after the correction
factor (RHS: distribution of its log) has been included.
LHS: error bars $\approx2$ size of the symbols.}
\vspace*{-0.6cm}
\end{figure}

An elementary yet suitable quantity for investigating the differences between
the theories is the static quark-antiquark potential, which is shown in
fig.$\,$2. The full and the quenched potential look familiar.
The ``topologically unquenched'' recipe generates a sample in between:
the raw potential (no correction factor being used) shows insufficient yet
nontrivial screening at long distances.
The log of the correction factor turns out to be sufficiently close to zero
to allow for a reconstruction of the potential as found with standard (full)
importance sampling.
In terms of CPU time, however, there is a big difference: In the present
example, the standard approach turned out to be more expensive by {\em two
orders of magnitude}.

\begin{figure}[t]
\epsfig{file=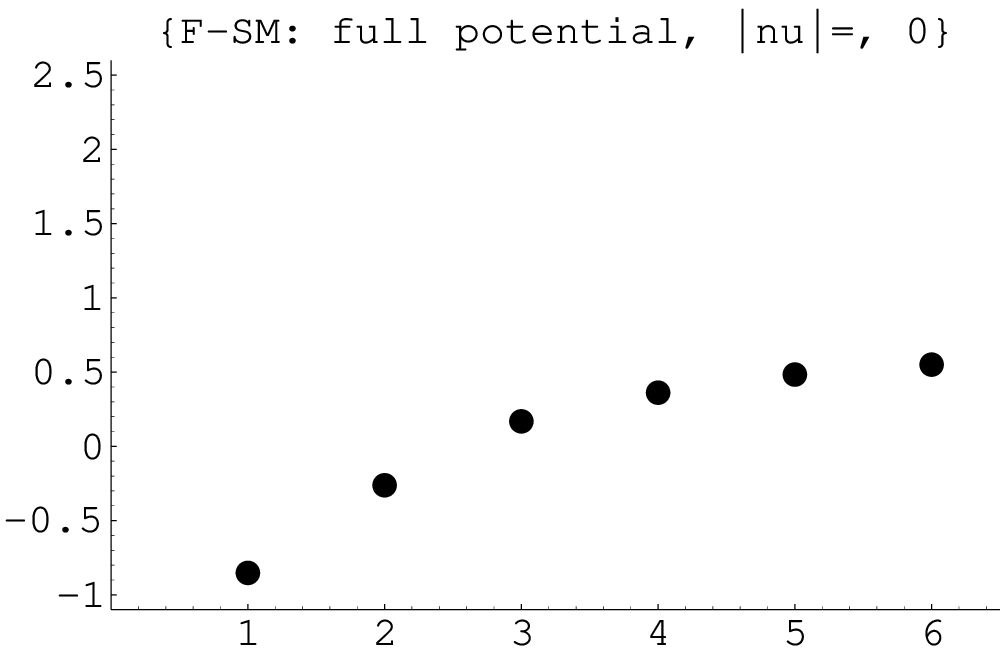,height=2.2cm,width=3.6cm,angle=0}
\hspace*{-0.1cm}
\epsfig{file=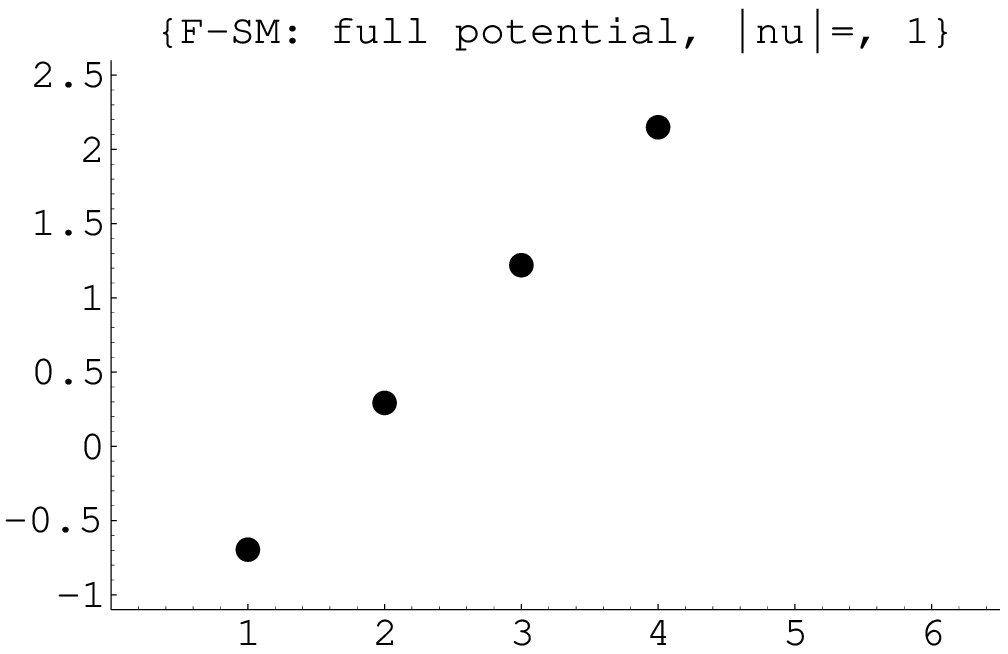,height=2.2cm,width=3.6cm,angle=0}
\vspace*{-0.8cm}
\caption{\sl Potential in the full theory, if only configurations with
$\nu\!=\!0$ or $\nu\!=\!\pm1$ are evaluated.}
\vspace*{-0.6cm}
\end{figure}

The physical reason why ``topologically unquenched'' importance sampling turns
out to be useful is elucidated by considering the hypothetical potentials
in the full theory, if only configurations in a certain topological sector
are taken into account.
As one can see from fig.$\,$3, these ``sectoral potentials'' agree with the
correct potential at short distances, but they deviate at larger separations.
Given the fact that the correct potential is, loosely speaking, a weighted
average of these sectoral potentials, it is clear that it is desirable
to preserve the relative weight of the different topological sectors in the full
theory --- even if importance sampling is done w.r.t. to a different measure.
Needless to say that this is exactly what the ``topological unquenching''
proposal is aiming at.

Issues which need more extensive coverage (e.g. potential sensitivity on
accidental misidentification of $\nu$, typical correction factors in larger
volumes, physical reason for the effect seen in fig. 3) are hoped to be
presented elsewhere.

\end{document}